\documentclass[12pt,epsf]{article}
\usepackage{graphicx}
\usepackage{amssymb}
\usepackage{amsmath}
\begin{document}

\def\a{\alpha}
\def\b{\beta}
\def\c{\varepsilon}
\def\d{\delta}
\def\e{\epsilon}
\def\f{\phi}
\def\g{\gamma}
\def\h{\theta}
\def\k{\kappa}
\def\l{\lambda}
\def\m{\mu}
\def\n{\nu}
\def\p{\psi}
\def\q{\partial}
\def\r{\rho}
\def\s{\sigma}
\def\t{\tau}
\def\u{\upsilon}
\def\v{\varphi}
\def\w{\omega}
\def\x{\xi}
\def\y{\eta}
\def\z{\zeta}
\def\D{\Delta}
\def\G{\Gamma}
\def\H{\Theta}
\def\L{\Lambda}
\def\F{\Phi}
\def\P{\Psi}
\def\S{\Sigma}

\def\o{\over}
\newcommand{\gsim}{ \mathop{}_{\textstyle \sim}^{\textstyle >} }
\newcommand{\lsim}{ \mathop{}_{\textstyle \sim}^{\textstyle <} }
\newcommand{\vev}[1]{ \left\langle {#1} \right\rangle }
\newcommand{\bra}[1]{ \langle {#1} | }
\newcommand{\ket}[1]{ | {#1} \rangle }
\newcommand{\EV}{ {\rm eV} }
\newcommand{\KEV}{ {\rm keV} }
\newcommand{\MEV}{ {\rm MeV} }
\newcommand{\GEV}{ {\rm GeV} }
\newcommand{\TEV}{ {\rm TeV} }
\def\diag{\mathop{\rm diag}\nolimits}
\def\Spin{\mathop{\rm Spin}}
\def\SO{\mathop{\rm SO}}
\def\O{\mathop{\rm O}}
\def\SU{\mathop{\rm SU}}
\def\U{\mathop{\rm U}}
\def\Sp{\mathop{\rm Sp}}
\def\SL{\mathop{\rm SL}}
\def\tr{\mathop{\rm tr}}

\newcommand{\beq}{\begin{equation}}   
\newcommand{\eeq}{\end{equation}}
\newcommand{\bea}{\begin{eqnarray}}   
\newcommand{\eea}{\end{eqnarray}}
\newcommand{\bear}{\begin{array}}  
\newcommand {\eear}{\end{array}}
\newcommand{\bef}{\begin{figure}}  
\newcommand {\eef}{\end{figure}}
\newcommand{\bec}{\begin{center}}  
\newcommand {\eec}{\end{center}}
\newcommand{\non}{\nonumber}  
\newcommand {\eqn}[1]{\beq {#1}\eeq}
\newcommand{\la}{\left\langle}  
\newcommand{\ra}{\right\rangle}
\newcommand{\ds}{\displaystyle}
\def\SEC#1{Sec.~\ref{#1}}
\def\FIG#1{Fig.~\ref{#1}}
\def\EQ#1{Eq.~(\ref{#1})}
\def\EQS#1{Eqs.~(\ref{#1})}
\def\GEV#1{10^{#1}{\rm\,GeV}}
\def\MEV#1{10^{#1}{\rm\,MeV}}
\def\KEV#1{10^{#1}{\rm\,keV}}
\def\lrf#1#2{ \left(\frac{#1}{#2}\right)}
\def\lrfp#1#2#3{ \left(\frac{#1}{#2} \right)^{#3}}


\baselineskip 0.7cm

\begin{titlepage}

\begin{flushright}
IPMU-10-0220  \\
\end{flushright}

\vskip 1.35cm
\begin{center}
{\large \bf
Strong dynamics at the Planck scale as a solution to the cosmological
moduli problem
}
\vskip 1.2cm
Fuminobu Takahashi$^a$
and 
Tsutomu T. Yanagida$^{a,b}$

\vskip 0.4cm

{\it $^a$Institute for the Physics and Mathematics of the Universe,
University of Tokyo, Kashiwa 277-8568, Japan}\\
{\it $^b$Department of Physics, University of Tokyo, Tokyo 113-0033, Japan}

\vskip 1.5cm

\abstract{ We argue that strong dynamics at the Planck scale can solve
  the cosmological moduli problem.  We discuss its implications for
  inflation models, and find that a certain type of multi-field
  inflation model is required for this mechanism to work, since
  otherwise it would lead to the serious $\eta$-problem.  Combined
  with the inflaton-induced gravitino problem, we show that a chaotic
  inflation with a discrete symmetry naturally avoids both
  problems. Interestingly, the focus point supersymmetry is predicted
  when this mechanism is applied to the Polonyi model.  }
\end{center}
\end{titlepage}

\setcounter{page}{2}

\section{Introduction}
In supergravity/superstring theories, there exist moduli fields,
collectively denoted by $Z$, which have flat potentials and obtain
masses from supersymmetry (SUSY) breaking.  In gravity mediation, for
instance, there is typically a pseudomodulus with a mass of order the
gravitino mass $m_{3/2}$ as in the Polonyi model.  During inflation
the minimum of the effective potential for such a modulus is
generically deviated from the low-energy minimum. After inflation, the
modulus starts to oscillate about the minimum with an amplitude of
order the Planck scale $M_p$, and soon dominates the energy density of
the Universe.  Since interactions are suppressed by the Planck scale,
the modulus lifetime is very long, leading to an onset of a
radiation-dominated Universe with a low temperature, typically below
MeV. Such a low temperature would dramatically alter the big bang
nucleosynthesis (BBN) predictions of light element abundances in
contradiction with observations.  This is the cosmological moduli
problem or the Polonyi
problem~\cite{Coughlan:1983ci,Goncharov:1984qm}.

Several solutions to the problem have been proposed so far. The
modulus abundance can be diluted if there is a huge entropy production
at late times by e.g. thermal
inflation~\cite{Lyth:1995ka,Randall:1994fr,Kawasaki:2004rx}, which
however also dilutes the pre-existing baryon
asymmetry~\cite{Stewart:1996ai}.  Another possible solution is to
suppose that the modulus mass is as heavy as $100$\,TeV, or heavier,
so that the modulus decays before BBN~\cite{Banks:1993en}. However,
the cosmological moduli problem still persists because decays of
moduli to gravitinos and gauginos are not suppressed~\cite{mig}.  If
the modulus mass is heavier than the gravitino, a pair of gravitinos
is generically produced by the modulus decay~\cite{mig,mig-c}. Those
gravitinos typically spoil the success of BBN or the lightest
supersymmetric particles (LSPs) produced by the gravitino decay exceed
the dark matter abundance, unless the gravitino is much heavier than
$O(100)$\,TeV.  On the other hand, the gravitino production from the
modulus decay can be kinematically forbidden if the modulus mass is
lighter than twice the gravitino mass. However, the branching fraction
of the modulus decay into SUSY particles is not suppressed~\cite{mig},
leading to an overproduction of the LSPs~\cite{Endo:2006ix}.\footnote{
In Refs.~\cite{Acharya:2008bk,Acharya:2009zt}, it was pointed out that
the Wino-like LSP of mass about $100$\,GeV can account for the present 
dark matter abundance if the modulus decay rate is enhanced by $O(10)$. 
If we adopt the cosmological constraints on the Wino mass~\cite{Hisano:2009rc,
Kanzaki:2009hf}, the modulus decay rate must be enhanced by $O(100)$ or greater.
We would like to thank Bobby Acharya,
Gordon Kane, Piyush Kumar, and Scott Watson for their comment on this issue.
}

In fact, there is a simple and elegant solution to the cosmological
moduli problem proposed long ago by Linde~\cite{Linde:1996cx}.  If the
inflaton and the modulus have a large quartic coupling,
\beq
{\cal L} \;=\; - \kappa \int d^4 \theta\,  \chi^\dag \chi Z^\dag Z
\label{quartic}
\eeq
with $\kappa = O(100)$, the amplitude of the modulus oscillations
becomes exponentially suppressed, where we adopt the Planck units,
$M_p = 1$.  Here $\chi$ denotes a chiral superfield which dominates
the energy density of the Universe when $Z$ starts to oscillate.  In
principle one can introduce such an enhanced operator by hand, however
the origin of such large coupling remains a puzzle.

In this paper we argue that the operator (\ref{quartic}) may result
from a strong dynamics at the Planck scale.  We will show that, using
the naive dimensional analysis (NDA)~\cite{NDA}, the operator
(\ref{quartic}) with a desired magnitude arises in the effective
theory below the dynamical scale, if both $\chi$ and $Z$ are involved
in the strong dynamics. Here the strong dynamics should not be
confused with any other interactions such as the usual QCD
interactions or those in the dynamical SUSY breaking, which become
strong at scales much lower than the Planck scale.  Throughout this
paper we do not specify the strong dynamics which is supposed to take
place at a scale close to the Planck scale. Such dynamics may
originate from the UV theory such as string
theory~\cite{Horava:1995qa}.

 We will  discuss the implications of such strong dynamics for the inflation models and find that a certain class of multi-field inflation model
is required for the mechanism to work, since otherwise it would make the $\eta$ problem much worse than usual.
In the following sections we mainly focus on the Polonyi problem as an explicit example, however, our main result can be straightforwardly applied to the generic moduli problem. 

The rest of the paper is organized as follows.
In Sec.~\ref{sec:2}, we describe a modified version of the Polonyi model 
with a strong dynamics at the Planck scale and show that the Polonyi problem can be solved in the model. 
We shall see in Sec.~\ref{sec:3} that the operator (\ref{quartic}) as well as many other ones 
are in general present, which restrict possible inflation models to a certain class of multi-field inflation models
in order not to make the $\eta$ problem worse. The last section is devoted to discussion and conclusions.

\section{A Polonyi model with strong dynamics}
\label{sec:2}
In this section we will show that strong dynamics at the Planck scale can solve the Polonyi 
problem. First we review the original Polonyi model and its cosmological problem, and
explain the Linde's proposal for a solution to the Polonyi problem. We will then see that 
an enhancement of the operator (\ref{quartic}) is
naturally realized if both  $\chi$ and $Z$ are strongly coupled at the Planck scale. 
Here and in what follows $\chi$ denotes a field that dominates the energy density of the Universe
when the $Z$ starts to oscillate after inflation. We will discuss whether the $\chi$ can represent
the slow-rolling inflaton in Sec.~\ref{sec:3}.

\subsection{The Polonyi problem and the Linde's solution}
We introduce a pseudomodulus $Z$ which is a singlet under any symmetries and therefore
has no special point in the field space. This property is required to give a soft mass to the SSM gauginos
from the following operator,
\beq
   \int d^2\theta\, Z\,W_\alpha W_\alpha ,
   \label{gaugino_mass}
\eeq
where $W_\alpha$ is a chiral superfield for the SM gauge multiplets.
The only scale associated with $Z$ is  considered to be
the Planck scale. These properties lead to the moduli (or Polonyi) problem as we shall see below.

The Polonyi model is given by
\bea
\label{K}
K &=& a_{10} Z + a_{01} Z^\dag+ |Z|^2 + \cdots \\
W &=&\Xi \left(1+c_1 Z + \cdots \right),
\label{W}
\eea
where $\Xi$ is a spectator field with a non-vanishing vacuum expectation value (VEV) $\langle \Xi\rangle = \mu$ 
and carries  an $R$ charge $2$, $a_{10} (= a_{01}^*)$ and $c_1$ are numerical coefficients of order unity, and
the kinetic term of $Z$ is set to be canonically normalized. 
Since $Z$ has an $R$-charge $0$, the K\"ahler potential is considered to be a generic function
of $Z$ and $Z^\dag$. 
The dots represent higher order
terms of $Z$ suppressed by the Planck scale, which are not relevant as long as we consider $|Z| < 1$.
Here and in what follows  we set the origin of $Z$ to be the potential minimum for simplicity.

The requirement of the vanishing cosmological constant relates $\mu$ and $c_1$ to the gravitino mass,
\beq
\mu\; = \; m_{3/2},~~~c_1 \mu\;=\;\sqrt{3} m_{3/2},
\eeq 
up to a phase factor. 
The $F$-term of $Z$ is  given by $F_Z =-\sqrt{3} m_{3/2}$. For a generic K\"ahler potential squarks,
sleptons and Higgs bosons  acquire the SUSY-breaking soft masses of $O(m_{3/2})$. The SSM gauginos
acquire a mass of the same order from the interaction (\ref{gaugino_mass}).

Examining the K\"ahler and super-potentials (\ref{K}) and (\ref{W}), one can see that the mass of $Z$ is of order the gravitino mass. 
In the early Universe, however, 
the effective potential of $Z$  is affected by the Planck suppressed interactions
with the inflaton sector.  Since there is no special point in the field space of $Z$, the potential minimum 
during inflation is generically deviated from the origin. 
When $H \sim m_{3/2}$, the $Z$ starts to oscillate about the 
origin with an amplitude of order the Planck scale, and soon dominates the energy density of the
Universe after reheating.\footnote{Around $|Z| \sim 1$, there will be SUSY vacua in general. Here we assume that 
$Z$ settles down to the SUSY breaking minimum at the origin.}
The couplings of $Z$ to the visible sector are suppressed by the Planck
scale, and the decay rate is roughly estimated by
\beq
\Gamma_Z \;\simeq\; \frac{c}{4 \pi} m_Z^3,
\label{dr}
\eeq
where $c$ is of order unity.
The decay temperature is given by
\bea T_Z &=& \lrfp{\pi^2 g_*}{90}{-\frac{1}{4}}
\sqrt{\Gamma_Z}\non\\ &\simeq& 0.002 \,{\rm \,MeV} \,c^{1/2}
\lrfp{m_{3/2}}{100{\rm\,GeV}}{3/2}, \eea where $g_*$ counts the
relativistic degrees of freedom, and $m_Z = m_{3/2}$ was used.  The
successful BBN requires $T_Z \gtrsim 5$\,MeV~\cite{Kawasaki:1999na}.
Thus, the onset of a radiation-dominated Universe is too late to be
consistent with observations, for the gravitino mass of order the weak
scale as in the gravity mediation. This is the notorious Polonyi
problem.

It was pointed out by Linde~\cite{Linde:1996cx} that if the $Z$ has an
enhanced coupling to $\chi$ as in (\ref{quartic}), the $Z$ follows a
time-dependent minimum and the effective amplitude of oscillations is
exponentially suppressed.  The enhanced quartic coupling
(\ref{quartic}) generates a mass of $Z$ as
\beq
m_Z \;=\; C H,
\eeq
where $C \sim \sqrt{\kappa}$.  The amplitude is suppressed by the
following factor~\cite{Linde:1996cx}
\beq
{\mathcal S} \;\simeq\; \frac{\sqrt{3 \pi}}{2} C^{\frac{3}{2}}  \exp{\left(- \frac{\pi C}{3}\right)},
\label{sup}
\eeq
where we have assumed an inflaton-matter domination at the onset of
oscillations of $Z$.\footnote{ This condition is necessary since too
  many gravitinos are produced if the reheating is already completed
  when $H \sim m_{3/2}$.}  The abundance of $Z$ is estimated as
\beq
Y_Z \; \sim \; \frac{1}{8} \frac{T_R}{m_Z} ({\cal S} Z_0 )^2,
\eeq
where $T_R$ denotes the reheating temperature and $Z_0 \sim 1$ is the
oscillation amplitude in the absence of the enhanced coupling.  It
depends on the mass of $Z$ how much its oscillation amplitude should
be suppressed to be consistent with BBN. For instance, the BBN
constraint reads, $Y_Z \lesssim O(10^{-16})$, for $m_Z =
100$\,GeV~\cite{Kawasaki:2004qu}.  Thus the suppression factor needs
to satisfy
\beq
{\cal S}  \;\lesssim\; 3 \times 10^{-10} Z_0^{-1} \lrfp{m_{3/2}}{100{\rm\,GeV}}{\frac{1}{2}}
\lrfp{T_R}{\GEV{6}}{-\frac{1}{2}} \lrfp{Y_Z^{\rm (BBN)}}{10^{-16}}{\frac{1}{2}},
\eeq
where $Y_Z^{\rm (BBN)}$ represents the BBN constraint on the abundance
of $Z$.  In order to achieve such suppression with the use of the
above mechanism, we need $C \sim 25$, or equivalently, $\kappa \sim
600$. For a heavier gravitino mass, the BBN bound is relaxed, and the
required value of $C$ becomes slightly smaller accordingly.

We comment on the validity of this mechanism. The essence of the
suppression is the adiabatic invariance. Namely, in the above Polonyi
model, the number density of $Z$ becomes a good adiabatic invariant in
the limit of $C \gg 1$, and that is why the coherent oscillations of
$Z$ is suppressed. The exponential factor in (\ref{sup}) reflects a
well-known fact that the variation of the adiabatic invariant is
exponentially suppressed.  Because of this, there is a limitation to
the potential of $Z$ where the mechanism applies. In the event that
the effective potential of $Z$ is extremely flat, the minimum of the
effective potential may change rapidly, and as a result the coherent
oscillations of $Z$ are induced even in the presence of the enhanced
coupling~\cite{Lukas:1996wx}.  This is expected to be the case in a
certain class of dynamical SUSY breaking models such as \cite{IYIT},
and the Polonyi problem still persists~\cite{Ibe:2006am}.  Thus, the
suppression mechanism applies only to the case in which the low-energy
potential of $Z$ is not much flatter than the quadratic potential in
the entire region where $Z$ moves, as in the Polonyi model.

The requisite for the Linde's solution is a quartic coupling with a
large coefficient. As long as we work in the low-energy effective
theory with a Planck-scale cut-off, such a large coefficient may look
a puzzle.  However, if there is a strong dynamics at the Planck scale,
the large coefficient may arise from the strong interaction. In fact,
using the NDA, we can roughly estimate the size of the coupling.  In
the next subsection we will see the coefficient obtained by NDA can
meet the requirement.

\subsection{Strong dynamics solves the Polonyi problem}
Now we show that the Polonyi problem can be solved if both $Z$ and
$\chi$ are strongly coupled at a scale $\Lambda$ close to the Planck
scale.  We will see that, using NDA, a coupling (\ref{quartic}) with a
desired magnitude arises from the strong dynamics.

Let us first describe a modified version of the Polonyi model.  We
assume that the Polonyi field $Z$ carries a vanishing $R$ charge as
usual, but it is assumed strongly coupled at a scale $\Lambda \simeq
O(1)$.  In addition to the $Z$ we introduce spectator fields carrying
an $R$ charge $2$, $\Xi$ and $\Xi^\prime$, which have a non-vanishing
VEV, $\langle \Xi\rangle = \mu$ and $\la \Xi^\prime \ra = \mu^\prime$.
Using NDA, we have a K\"ahler potential and a superpotential
\bea
\label{Ks}
K &\approx &\frac{\Lambda^2}{16 \pi^2}\,\sum_{i,j=0}  a_{ij} \lrfp{ 4 \pi Z}{\Lambda}{i} 
 \lrfp{ 4 \pi Z^\dag}{\Lambda}{j}\non\\
&=&\frac{\Lambda}{4 \pi} \left(a_{10}  Z + a_{01} Z^\dag \right) + 
\left(a_{20} Z^2 + a_{02} Z^{\dag 2}  \right) + |Z|^2 +  \cdots
\\
W &\approx &\Xi^\prime+ \Xi \cdot   \sum_{i=0}  c_i \lrfp{ 4 \pi Z}{\Lambda}{i} \non\\
&=&(\Xi^\prime+\Xi)+\Xi \left( c_1 \frac{4 \pi Z}{\Lambda} +  c_2 \lrfp{4 \pi Z}{\Lambda}{2}+
c_3  \lrfp{4 \pi Z}{\Lambda}{3} + \cdots
\right)
\label{Ws}
\eea
where $\Xi^\prime$ is assumed to be decoupled from $Z$, $a_{ij}
(=a_{ji}^*)$ and $c_i$ are numerical coefficients of order unity, and
we have normalized $a_{11} = 1$ and $c_0 = 1$.  We have included a
factor of $1/16 \pi^2$, which usually appears in NDA, in the
definition of $\Xi$ and $\Xi^\prime$.  For the above NDA to be valid,
the value of the Polonyi field is constrained as
\beq
|Z| \;\lesssim\; \frac{\Lambda}{4 \pi}.
\label{upZ}
\eeq
There is in general a SUSY breaking meta-stable vacuum at $|Z| <
\Lambda/4\pi$, while there are SUSY preserving vacua at $|Z| \sim
\Lambda/4\pi$.  In supergravity, the scalar potential is considered to
increase exponentially for $|Z| > \Lambda/4\pi$ because of an
exponential pre-factor, $e^K$, in the scalar potential. As we shall
see later, this constraint on the variation of $Z$ will be important
when we discuss the implications for inflation models.  We will set
the SUSY breaking minimum to be at the origin, which places a certain
relation among the coefficients $a_{ij}$ and $c_i$, but the following
argument is not affected.  The requirement of the vanishing
cosmological constant is satisfied if
\beq
\mu + \mu^\prime = m_{3/2},~~~\frac{4 \pi c_1 \mu}{\Lambda} = \sqrt{3} m_{3/2}.
\eeq
While the $F$-term of $Z$ is given by $F_Z = - \sqrt{3} m_{3/2}$ as
before, the soft masses of SUSY particles are modified. Assuming that
the SSM particles are not involved in the strong dynamics at the
Planck scale, the scalars acquire a soft SUSY breaking mass of order
the gravitino mass for a generic K\"ahler potential. On the other
hand, the gaugino mass arises from the following operator instead of
(\ref{gaugino_mass}),
\beq
   \int d^2\theta\, \frac{Z}{4 \pi \Lambda}\,W_\alpha W_\alpha.
   \label{gaugino_mass2}
\eeq
The gaugino mass is of order $O(m_{3/2}/4\pi\Lambda)$, an order of
magnitude lighter than the scalar mass. The gaugino masses are
typically of $O(100)$\,GeV, while the scalar masses are several
TeV. Thus, the SUSY mass spectrum is that in the focus point
region~\cite{Feng:1999hg}, which has phenomenological virtues. It is
interesting that the focus point SUSY naturally appears from the
strong dynamics at the Planck scale.

The mass of $Z$ about the origin mainly arises from the quartic
coupling in the K\"ahler potential,
\beq
K \;\supset\; a_{22} \lrfp{4 \pi}{\Lambda}{2} |Z|^4,
\label{z4}
\eeq
leading to
\beq
m_Z \;\sim \; \frac{4 \pi}{\Lambda} m_{3/2}.
\label{mznda}
\eeq
Here we have assumed that the sign of $a_{22}$ is negative for the
stability of the SUSY breaking vacuum.

If the $\chi$ is not involved in the strong dynamics, the coefficient
$\kappa$ of (\ref{quartic}) is expected to be of order unity. Then,
the Polonyi field $Z$ is generically away from the origin during
inflation, and starts to oscillate after inflation when the Hubble
parameter becomes comparable to its mass. Since it has a large initial
amplitude $Z_0 \sim \Lambda/4 \pi$ and it decays only through
interactions suppressed by $\Lambda$, the Polonyi problem still
persists.  The decay rate of $Z$ to the SSM particles is modified from
(\ref{dr})
\beq
\Gamma(Z \rightarrow {\rm SSM~particles}) \;\sim\; \frac{c}{4 \pi} \frac{m_Z^3}{(4 \pi \Lambda)^2} \sim  \frac{c \, m_{3/2}^3}{\Lambda^5},
\label{Z2SM}
\eeq
where (\ref{mznda}) is used in the last equality. One may expect that
the decay temperature becomes higher than $5$\,MeV, since the mass of
$Z$ is larger than before.  Actually, however, the main decay mode of
$Z$ is to a pair of gravitinos through (\ref{z4}), and the decay rate
is given by~\cite{mig}
\beq
\Gamma(Z \rightarrow 2 \psi_{3/2}) \;\simeq\; \frac{1}{96 \pi} \frac{m_Z^5}{m_{3/2}^2}
\eeq
which is several orders of magnitudes larger than (\ref{Z2SM}). Thus,
the Universe will be dominated by the gravitinos produced by decay of
$Z$. The gravitino of mass $\simeq O(1)$\,TeV decays during BBN and it
becomes inconsistent with observations.  In order to avoid the BBN
constraint, the suppression factor ${\cal S}$ for the oscillation
amplitude $Z_0$ must satisfy
\beq
{\cal S}  \;\lesssim\; 3 \times 10^{-8}  \lrfp{m_{3/2}}{1\,{\rm\,TeV}}{\frac{1}{2}}
\lrfp{T_R}{\GEV{6}}{-\frac{1}{2}} \lrfp{\Lambda}{M_p}{-\frac{3}{2}}\lrfp{Y_Z^{\rm (BBN)}}{10^{-16}}{\frac{1}{2}}.
\label{sup2}
\eeq
Thus the Polonyi problem is not solved if only $Z$ is strongly coupled
at the Planck scale.

Now let us assume that $\chi$ is also involved in the strong
dynamics. Using NDA, we expect that there is a quartic coupling
(\ref{quartic}) with
\beq
\kappa \sim \frac{16 \pi^2}{\Lambda^2}.
\eeq
Using (\ref{sup}), the suppression of $O(10^{-8})$ is achieved for $C
\sim 22$ or $\kappa \sim 500$.  If we take $\Lambda \sim 0.6$, this
condition is satisfied.  Thus, the strong dynamics at the Planck scale
indeed solves the Polonyi problem.  It is straightforward to check
that one arrives at the essentially same conclusion for the generic
moduli problem, except for the prediction of the SUSY mass spectrum.

\section{Implications for inflation models}
\label{sec:3}
We have seen that the strong dynamics can solve the moduli problem.
Here we discuss its implications for the inflation models.

The crucial assumption was that both $\chi$ and $Z$ are strongly
coupled at the Planck scale.  For the moment we assume that $\chi$ and
$Z$ are the only particles in the low energy effective theory, which
are involved in the strong dynamics.  The NDA provides us with a
prescription to estimate the size of interactions allowed by the
symmetry. While $Z$ must be a singlet under any symmetries to generate
gaugino masses through (\ref{gaugino_mass2}), we assume a non-trivial
charge on $\chi$ to forbid unwanted couplings like $W = \chi Z$. This
is because such an operator induces a large SUSY mass for $Z$ and $Z$
should no longer be treated as a modulus field.  In general, however,
we expect that there are unsuppressed interactions involving
$|\chi|^2$, $Z$ and $Z^\dag$ in the K\"ahler potential.

Since $\chi$ gives a main contribution to the energy density of the
Universe at the onset of the modulus oscillations, it is natural to
expect that $\chi$ is a part of the inflation sector.  Let us first
consider a single-field inflation model in which $\chi$ is to be
identified with the inflaton.  One can easily see, however, that the
$\chi$ cannot be responsible for the slow-rolling inflaton, because it
would have a too large mass from the following operator~\footnote{ One
  exception is the case in which $\chi$ has a shift
  symmetry~\cite{Kawasaki:2000yn}.  },
\beq
\int d^4 \theta \frac{16 \pi^2}{\Lambda^2} |\chi|^4.
\label{quartic-inf}
\eeq
The mass of $\chi$ is then 
\beq
m_\chi \;\simeq\; O(10) H_{\rm inf},
\eeq
where $H_{\rm inf}$ represents the Hubble parameter during
inflation. Thus, the $\eta$-problem becomes worse than usual, and the
slow-roll inflation does not occur unless the above coupling
(\ref{quartic-inf}) is suppressed somehow or there is an accidental
cancellation between (\ref{quartic-inf}) and other contributions.  So
we conclude that our mechanism does not fit with the single-field
inflation model in which $\chi$ is identified with the inflaton.

In principle, there could be another weakly-coupled scalar field
responsible for the slow-rolling inflaton.  In this case, the
$\eta$-problem does not necessarily become worse. So we are led to
consider an inflation model in which there are multiple fields, and
one (or some) of the fields is involved in the strong dynamics while
the slow-rolling inflaton is weakly coupled.  To be explicit, let us
focus on a class of two-field inflation models with the following
superpotential,
\beq
W\;=\; X f(\phi),
\label{2inf}
\eeq
where $X$ and $\phi$ are chiral superfields, and $f(\phi)$ is some function of $\phi$. 
We assume that $X$ and $\phi$ have  $R$-charge $2$ and $0$, respectively.
During inflation, the $X$ has a non-vanishing $F$-term, $-F_X^* = f(\phi)$, which drives inflation.
Thus, the value of $f(\phi)$ during inflation determines the inflation scale;
\beq
|f(\phi)_{\rm inf}| \;\simeq\; \sqrt{3} H_{\rm inf} ~~~({\rm during~~inflation}),
\label{finf}
\eeq
where $\phi_{\rm inf}$ represents its typical value during inflation.
After inflation both $X$ and $\phi$ are assumed to settle down to the SUSY minimum. In particular, $f(\phi)$
must vanish at the minimum;
\beq
f(\phi_{\rm min}) = 0.
\label{fmin}
\eeq
It has been known that the above class of models cover many inflation
models~\cite{Kawasaki:2006gs}, such as hybrid
inflation~\cite{Copeland:1994vg}, smooth hybrid
inflation~\cite{Lazarides:1995vr}, multi-field new
inflation~\cite{Asaka:1999jb}, and chaotic
inflation~\cite{Kawasaki:2000yn} as well as its
variants~\cite{Takahashi:2010ky,Kallosh:2010ug}.  In the hybrid and
smooth hybrid inflation $X$ is the slow-rolling inflaton, while $\phi$
plays a role of the inflaton in the other models.  In order to avoid
the above-mentioned $\eta$-problem, therefore, the $\chi$ needs to be
identified with $\phi$ $(X)$ in the former (latter) case.

In fact, however, there is generically a fine-tuning problem in the
case where $X$ is the inflaton and $\phi$ is a strongly coupled
field. Since both $\phi$ and $Z$ are involved in the strong dynamics
and $Z$ is a singlet, we expect that there is a coupling
\beq
\int d^2 \theta \,X \tilde{f}(\phi) \lrf{4\pi Z}{\Lambda},
\label{int2}
\eeq
where $\tilde{f}(\phi)$ is a function of $\phi$, which does not
necessarily coincide with $f(\phi)$, but is expected to contain the
same $\phi$-dependent interactions in $f(\phi)$.  For instance, if
$f(\phi)$ contains a $\phi^2$-term, $\tilde{f}(\phi)$ generically
contains $\phi^2$ with a coefficient of the same order.  Then, one can
see from (\ref{finf}) and (\ref{fmin}) that $\tilde{f}(\phi)$ should
change its value by $O(H_{\rm inf})$ unless there is an accidental
cancellation,
\beq
\left|\tilde{f}(\phi_{\rm inf}) - \tilde{f}(\phi_{\rm min})\right| = O(H_{\rm inf}).
\eeq
However, this  causes the following problem.
First of all, $\tilde{f}(\phi_{\rm min})$ must be vanishingly small since otherwise the $Z$ would acquire a large
SUSY mass and the SUSY would not be broken.
Therefore we have $|\tilde{f}(\phi_{\rm inf})| = O(H_{\rm inf})$.
The interaction (\ref{int2})  then generates a mass of $O(10)H_{\rm inf}$ for the inflaton $X$,
and the slow-roll inflation does not occur. So we need to fine-tune the function $|\tilde{f}(\phi_{\rm inf})| \ll H_{\rm inf}$.
Thus, this class of models (e.g. the hybrid and smooth hybrid inflation)
is plagued with  the serious $\eta$-problem.\footnote{In fact, the inflaton-induced gravitino problem is also 
more serious than usual.}

On the other hand, if $\phi$ is the inflaton and $X$ is identified
with $\chi$, there is similarly an interaction like (\ref{int2}) with
$|\tilde{f}(\phi_{\rm inf})| = O(H_{\rm inf})$, which gives a mass of
$O(10)H_{\rm inf}$ to the $X$. However, the inflaton potential
receives only a slight modification, and it does not make the
$\eta$-problem worse than usual. The scalar potential during inflation
is given by
\beq
V\;\sim\; \left|f(\phi) + \tilde{f}(\phi) \frac{4 \pi Z_0}{\Lambda}\right|^2,
\eeq
where we assume that $X$ and $Z$ are stabilized at the origin and at
$Z_0$, respectively.  Since $\tilde{f}(\phi)$ has the same functional
dependence on $\phi$ as $f(\phi)$, the inflaton dynamics is not
modified. We emphasize here that it is crucial for the inflation that
$Z$ is bounded above as $Z_0 \lesssim \Lambda/4\pi$~\footnote{ See
  discussion below (\ref{upZ}).  }, since otherwise $Z$ would be
stabilized at the SUSY minimum and inflation would not occur.

To see this explicitly, let us consider the chaotic inflation with a
shift symmetry~\cite{Kawasaki:2000yn}.  The K\"ahler and
super-potentials are
\bea
K &=& |X|^2+\frac{1}{2} (\phi+\phi^\dag)^2 + \cdots,\\
W&=& m X \phi,
\eea
where $\phi$ has a shift symmetry: $\phi \rightarrow \phi + i \alpha$
with $\alpha \in {\bf R}$, and $m (\sim 10^{-5})$ represents the
breaking of the shift symmetry.  As long as $Z$ is stabilized at $Z_0
\sim \Lambda/4\pi$, including the coupling
\beq
\int d^2\theta\, m X \phi \lrf{4 \pi Z}{\Lambda}
\eeq
slightly shifts the inflaton mass $m$, but the form of the inflaton
potential, $m^2 |\phi|^2$, is not changed.  In particular, such an
interaction does not lead to the $\eta$-problem.\footnote{ There is no
  $\eta$-problem in the chaotic inflation of
  Ref.~\cite{Kawasaki:2000yn} because of the shift symmetry.  In the
  case of the multi-field new inflation, the light inflaton mass is
  realized by tuning the interactions in the K\"ahler potential. The
  amount of the tuning remains the same even if we add an interaction
  like (\ref{int2}).}

To summarize, our mechanism is consistent with a class of two-field
inflation models (\ref{2inf}), in which $\phi$ plays the role of
inflaton and $X$ is identified with the strongly coupled field $\chi$.
The inflation models satisfying this condition contain a multi-field
new inflation and a chaotic inflation.\footnote{Actually it is
  possible to assume that both $\phi$ and $X$ are strongly coupled in
  the chaotic inflation, because the $\eta$-problem is absent due to
  the shift symmetry.}

Lastly let us mention the inflaton-induced-gravitino
problem~\cite{Kawasaki:2006gs,Endo:2007ih}.  We identify $X$ and
$\phi$ with $\chi$ and the inflaton, respectively.  We expect that
there is the following interaction,
\beq
\int d^4\theta\,  |\phi|^2 Z Z + {\rm h.c.}.
\eeq
Noting that the fermionic component of $Z$ is a goldstino, this
operator induces the decay into a pair of the
gravitinos~\cite{Kawasaki:2006gs,mig-c},
\beq
\Gamma(\phi \rightarrow 2 \psi_{3/2}) \;\simeq\; 
\frac{1}{8 \pi} \la \phi \ra^2 m_\phi^3,
\eeq
where $\la \phi \ra$ denotes the VEV of $\phi$.\footnote{ Precisely
  speaking, the effective partial decay rate into the gravitinos
  should be multiplied with $1/2$, because $\phi$ and $X$ are
  maximally mixed in the vacuum in the models given by
  (\ref{2inf})~\cite{Kawasaki:2006gs,Endo:2007ih}.  }

The constraint on the gravitino production from the inflaton decay was
studied in detail for a variety of inflation models in
Refs.~\cite{Kawasaki:2006gs,Endo:2007ih}.  Applying the results of
Refs.~\cite{Kawasaki:2006gs,Endo:2007ih} to our case, we conclude that
gravitinos are overproduced in the multi-new inflation model unless
the hadronic branching fraction of the gravitino is as small as
$10^{-3}$.  What is peculiar to the chaotic inflation is that one can
assign a $Z_2$ symmetry on $\phi$ and
$X$~\cite{Kawasaki:2000yn,Takahashi:2010ky}, forbidding the VEV,
i.e. $\la \phi \ra = 0$. Thus there is no non-thermal gravitino
problem in the chaotic inflation with the $Z_2$ symmetry.

To summarize, the strong dynamics at the Planck scale is consistent
with a slow-roll inflation for a certain type of multi-field
inflation, in which the strongly coupled field, $\chi$, corresponds to
a field other than the slow-rolling inflaton.  Combined with the
inflaton-induced gravitino problem, it is the chaotic inflation with
the discrete symmetry that leads to a successful cosmology.

\section{Discussion and Conclusions}
\label{sec:4}

In the previous section we have discussed implications of the strong
dynamics for inflation models.  In principle, the $\chi$, which has an
enhanced quartic coupling (\ref{quartic}) with a modulus $Z$, could
have no relation with the inflation sector. For instance, the inflaton
may decay fast and the energy density of $\chi$ may eventually come to
dominate the energy density of the Universe before $Z$ starts to
oscillate. Such situation may be realized in a curvaton
scenario~\cite{curvaton} where the $\chi$ plays the role of the
curvaton.  In this case, we can avoid some of the constraints on the
inflation models discussed in Sec.~\ref{sec:3}. However, in the event
that $Z$ is the Polonyi field responsible for the SUSY breaking, the
non-thermal gravitino production still takes place as long as $\chi$
reheats the Universe. In order not to induce too strong coupling
between the SSM sector and the $Z$, the couplings of $\chi$ to the SSM
sector should be suppressed by the Planck scale. Then, the branching
fraction of the gravitino production is typically of $O(\la \chi
\ra^2)$, and too many gravitinos are produced by the $\chi$ decay
unless the VEV of $\chi$ is suppressed by some symmetry. In summary,
even if $\chi$ is not a part of the inflation sector, a successful
cosmology requires a sufficiently small $\la \chi \ra$ in order to
suppress the gravitino abundance, which suggests that $\chi$ is
charged under some symmetry.

The Polonyi problem arises from the presence of a singlet $Z$ with a
non-vanishing $F$-term, which is required in the gravity mediation in
order to generate the SSM gaugino masses. On the other hand, in gauge
and anomaly mediation models~\cite{GMSB,AMSB}, we can assign a
non-trivial charge on a SUSY breaking field.  Then the SUSY breaking
field sits at the symmetry point during inflation and no sizable
coherent oscillations are induced.  Thus there is no Polonyi
problem. Of course, if there is a modulus field, its coherent
oscillations lead to the cosmological moduli problem, which can be
solved by the strong dynamics at the Planck scale discussed in this
paper.

We have assumed that only $\chi$ and $Z$ are involved in the strong
dynamics. Let us discuss to what extent we can extend this
assumption. First of all, suppose that the inflaton $\phi$ is also
strongly coupled. This leads to the serious $\eta$-problem in general,
except for the chaotic inflation in which $\phi$ has a shift symmetry
and its light mass is protected by the symmetry.  Also, if $Z$ is the
Polonyi field, the inflaton-induced gravitino problem becomes
worse. Again, the chaotic inflation avoids the problem if a discrete
symmetry is assigned on the inflaton.  Next let us consider the case
that all the particles including the SSM particles are also strongly
coupled at the Planck scale. The couplings of the Polonyi field $Z$
with the SSM sector are modified, and the scalar and gaugino masses
become of $O(4 \pi m_{3/2}/\Lambda)$. Thus the gravitino is the LSP
with a mass an order-of-magnitude lighter than the sfermion and gaugino masses, i.e.,
$m_{3/2} = O(10)\,$GeV up to $100$\,GeV. Such a mass spectrum is
favored from the thermal leptogenesis scenario~\cite{Fukugita:1986hr}
and the gravitino problem. The BBN bound on the decay of the
next-to-lightest SUSY particle can be avoided if the $R$-parity is
violated~\cite{Buchmuller:2007ui,Kuriyama:2008pv}.  
We also note that the cosmological problem associated with the saxion, the scalar
partner of the axion in the Peccei-Quinn
mechanism~\cite{Peccei:1977hh}, can be solved by the same mechanism.

We notice that a theory where all the fields except for
the gravity multiplets are involved in the strong dynamics may be interpreted as
a result of a lower cut-off theory, $M_{\rm CUT}  \ll M_p$. Here we
replace $\Lambda/4\pi$ with the cut-off scale $M_{\rm CUT}$. Similarly to the case of the strong coupling, 
when combined with the thermal leptogenesis scenario, the theory implies (i) the gravitino
dark matter with a mass of $\sim 100$\,GeV, and the squark, slepton and gaugino masses
of $O(1)$\,TeV, or greater; (ii) the R-parity breaking; (iii) an inflation model with a shift symmetry which
solves the $\eta$ problem~\cite{TY}.

In this paper we have argued that the solution to the moduli problem
proposed by Linde~\cite{Linde:1996cx} is naturally realized if there
is a strong dynamics at a scale close to the Planck scale, and if both
$\chi$ and $Z$ are involved in the dynamics. We have shown that an
operator like (\ref{quartic}) arises based on the NDA, whose
coefficient can be large enough to solve the moduli problem.  We have
considered implications for the inflation models, and found that a
class of two-field inflation models including the multi-field new
inflation as well as the chaotic inflation (but not the hybrid
inflation) is consistent with the strong dynamics. In the event that
the $Z$ is the Polonyi field responsible for the SUSY breaking, the
inflaton-induced gravitino problem tightly constrains the inflation
parameter space. However the problem can be avoided in the chaotic
inflation with a discrete symmetry.  We have also pointed out that the
predicted SUSY mass spectrum is that in the focus point region.  In
the absence of the Polonyi problem, the observed dark matter abundance
can be explained by the thermal relic of the lightest neutralino LSP
with a sizable mixture of the Higgsino component.

\section*{Acknowledgment}

FT thanks K.~Nakayama for useful discussion.
This work was supported by the Grant-in-Aid for Scientific Research on 
Innovative Areas (No. 21111006) [FT],  Scientific Research (A)
(No. 22244030 [FT] and 22244021 [TTY]), and JSPS Grant-in-Aid for Young Scientists (B) (No. 21740160) [FT]. 
This work was also supported by World Premier
International Center Initiative (WPI Program), MEXT, Japan.


\end{document}